\documentclass[aps,prl,floatfix,twocolumn,groupedaddress,showpacs,longbibliography]{revtex4-1}

\usepackage[utf8]{inputenc}
\usepackage[english]{babel}
\usepackage{amsmath,amssymb,graphicx,enumerate,bm}

\usepackage{color}

\newcommand\beq{\begin{equation}}
\newcommand\eeq{\end{equation}}
\newcommand\beqa{\begin{eqnarray}}
\newcommand\eeqa{\end{eqnarray}}
\newcommand{\nn}{\nonumber\\}
\def\bal#1\eal{\begin{align}#1\end{align}}
\newcommand{\re}{{\text{ref}}}
\newcommand{\st}{{\text{st}}}
\newcommand{\hcs}{{\text{HCS}}}

\begin{document}

\title{When the Hotter Cools More Quickly: Mpemba Effect in Granular Fluids}
\author{Antonio Lasanta$^{1,2}$}
\author{Francisco Vega Reyes$^{2}$}
\author{Antonio Prados$^{3}$}
\author{Andr\'es Santos$^{2}$}
\affiliation{
$^1$Gregorio Mill\'an Institute of Fluid Dynamics,
Nanoscience and Industrial Mathematics,
Department of Materials Science and Engineering and Chemical Engineering,
Universidad Carlos III de Madrid, 28911 Legan\'es, Spain\\
$^2$Departamento de F\'{\i}sica and Instituto de Computaci\'on Cient\'{\i}fica Avanzada (ICCAEx), Universidad de Extremadura, 06006 Badajoz, Spain \\
$^3$F\'{\i}sica Te\'orica, Universidad de Sevilla, Apartado de Correos 1065, 41080
Sevilla, Spain}

\date{\today}

\begin{abstract}
  Under certain conditions, two samples of fluid at  different initial
  temperatures present a counterintuitive behavior known as the
  Mpemba effect: it is the hotter system that cools sooner.  Here, we
  show that the Mpemba effect is present in granular fluids, both in uniformly heated and in freely cooling systems. In both
  cases, the system remains homogeneous, and no phase transition is
  present. Analytical quantitative predictions are given for how
  differently the system must be initially prepared to observe the
  Mpemba effect, the theoretical predictions being confirmed by both
  molecular dynamics and Monte Carlo simulations. Possible
  implications of our analysis for other systems are also discussed.
\end{abstract}

\maketitle

Let us consider two identical beakers of water, initially at two different
temperatures, put in contact with a thermal reservoir at
subzero (on the Celsius scale) temperature. While one may intuitively
expect that the initially cooler sample would freeze first, it has
been observed that this is not always the case
\cite{mpemba_cool?_1969}. This paradoxical behavior named the Mpemba
effect (ME) has been known since antiquity and discussed by
philosophers like Aristotle, Roger Bacon, Francis Bacon, and Descartes
\cite{jeng_mpemba_2006,B11}. Nevertheless, physicists only started to
analyze it in the second part of the past century, mainly in popular
science or education journals
\cite{jeng_mpemba_2006,B11,mpemba_cool?_1969,kell_freezing_1969,F71,D71,F74,G74,walker_hot_1977,O79,F79,K80,H81,A95,K96,CK06,katz_when_2009,WCVN11,BT12,S15,BT15,R15,IC16}.

There is no consensus on the underlying physical mechanisms that bring
about the ME. Specifically, water evaporation
\cite{kell_freezing_1969,F71,walker_hot_1977,VM11}, differences in
the gas composition of water \cite{F79,WOB88,katz_when_2009}, natural convection
\cite{D71,M96,IC16}, or the influence of supercooling, either alone
\cite{A95,zhang_hydrogen-bond_2014} or combined with other causes
\cite{esposito_mpemba_2008,VM12,VK15,jin_mechanisms_2015}, have been claimed to have an impact
on the ME. Conversely, the own existence of the ME in water has been
recently put in question \cite{BL16}. Notwithstanding, Mpemba-like
effects have also been observed in different physical systems, such as
carbon nanotube resonators \cite{greaney_mpemba-like_2011} or
clathrate hydrates \cite{AKKL16}.

The ME requires the evolution equation for the temperature to involve
other variables, which may facilitate or hinder the temperature
relaxation rate. The initial values of those additional variables
depend on the way the system has been prepared, i.e., ``aged,'' before
starting the relaxation process. Typically, aging and memory effects
are associated with slowly evolving systems with a complex energy
landscape, such as glassy
\cite{kovacs_isobaric_1979,bouchaud_weak_1992,prados_aging_1997,bonilla_aging_1998,berthier_geometrical_2002,mossa_crossover_2004,
  aquino_memory_2008,prados_kovacs_2010,ruiz-garcia_kovacs_2014} or
dense granular systems
\cite{nicodemi_aging_1999,josserand_memory_2000,brey_linear_2001}. However,
these effects have also been observed in simpler systems, like
granular gases
\cite{ahmad_velocity_2007,brey_scaling_2007,prados_kovacs-like_2014,trizac_memory_2014}
or, very recently, crumpled thin sheets and elastic foams
\cite{lahini_nonmonotonic_2017}.

In a general physical system, the study of the ME implies finding
those additional variables that control the temperature relaxation and
determining how different they have to be initially in order to
facilitate its emergence.  In order to quantify the effect with the
tools of nonequilibrium statistical mechanics, a precise definition
thereof is mandatory. An option is to look at the relaxation time to
the final temperature as a function of the initial temperature
\cite{mpemba_cool?_1969,kell_freezing_1969,F71,F79,walker_hot_1977,jeng_mpemba_2006,VM11,BL16,AKKL16}. Alternatively,
one can analyze the relaxation curves of the temperature: if the curve
for the initially hotter system crosses that of the initially cooler
one and remains below it for longer times, the ME is present
\cite{walker_hot_1977,esposito_mpemba_2008,B11,greaney_mpemba-like_2011,WCVN11,VM12,zhang_hydrogen-bond_2014,S15,VK15}.

In this Letter, we combine both alternatives above and investigate the
ME in a prototypical case of intrinsically out-of-equilibrium system:
a granular fluid \cite{H83,G03,PL01,AT09}, i.e., a (dilute or
moderately dense) set of mesoscopic particles that do not preserve
energy upon collision. As a consequence, the mean kinetic energy, or
{granular temperature} $T(t)$ \cite{G03}, decays monotonically in time unless an
external energy input is applied. The simplicity of the granular fluid
makes it an ideal benchmark for other, more complex, nonequilibrium
systems. We analyze the time evolution of the granular fluid starting from different initial preparations  and quantitatively investigate how the ME
appears. This is done for both the homogeneous heated and freely
cooling cases.

Our granular fluid is composed of smooth inelastic
hard spheres. Therefore, the component of the relative velocity along
the line joining the centers of the two colliding particles is
reversed and shrunk by a constant factor $\alpha$ \cite{H83},  the
so-called {coefficient of normal restitution}. In addition, the
particles are assumed to be subject to random forces in the form of a white-noise
thermostat with variance $m^{2}\xi^{2}$, where $m$ is the mass of a
particle. Thus, the velocity distribution function (VDF)
$f(\bm{v},t)$ obeys an Enskog-Fokker-Planck kinetic equation
\cite{vNE98,MS00,GMT12}.

The granular temperature and the excess kurtosis (or second Sonine
coefficient) are defined as
$ T(t)=\frac{m}{3} \langle v^2 (t) \rangle \equiv \frac{m}{3n} \int d\bm{v} \, v^2 f(\bm{v},t)$ and $a_{2}=\frac{3}{5}\langle v^{4}\rangle/\langle v^{2}\rangle^{2}-1$, respectively,
where $n= \int d \bm{v} f(\bm{v},t)$ is the number density.
From the kinetic equation for the VDF, one readily finds  \cite{vNE98}
\begin{subequations}\label{evol-eqs}
\bal
\label{evol-eqs-a}
\frac{dT}{dt}=&-\frac{2\kappa}{3}\left(\mu_{2}T^{3/2}-\chi\right),\\
\frac{d\ln(1+a_{2})}{dt}=&\frac{4\kappa}{3T} \left(\mu_{2}T^{3/2}-\chi-\frac{\frac{1}{5}\mu_{4}T^{3/2}-\chi}{1+a_{2}}\right),
\eal
\end{subequations}
where $\kappa\equiv 2ng(n)\sigma^{2}\sqrt{\pi/m}$, $\sigma$ and $g(n)$
are the sphere diameter and the pair correlation function at contact
\cite{CS69}, respectively, $\chi\equiv\frac{3m}{2\kappa}\xi^{2}$, and
$\mu_{2}$ and $\mu_{4}$ are dimensionless collisional rates.

Note that Eqs.\ \eqref{evol-eqs} are formally exact, but (a)  $T$ and  $a_2$ are coupled, and (b) the equations are not closed in those two variables since $\mu_n$ are functionals of the whole VDF. However,
if inelasticity is not too large, the nonlinear contributions of $a_{2}$ and the complete contributions of higher order cumulants  can be neglected. This is the so-called first Sonine approximation \cite{vNE98,SM09}, which yields
 $\mu_{n}\simeq\mu_{n}^{(0)}+\mu_{n}^{(1)} a_{2}$, with
$\mu_2^{(0)}=1-\alpha^2$, $\mu_2^{(1)}=\frac{3}{16}\mu_2^{(0)}$,
$\mu_4^{(0)}=\left(\frac{9}{2}+\alpha^2\right)\mu_2^{(0)}$,
$\mu_4^{(1)}=
(1+\alpha)\left[2+\frac{3}{32}(69+10\alpha^2)(1-\alpha)\right]$.

Using the first Sonine approximation above in Eqs.\ \eqref{evol-eqs}, they become a closed set, but the $T$-$a_2$ coupling still remains.
Taking into account this coupling, and since $\mu_2$ is an increasing function of $a_2$, it turns out that the relaxation of the granular temperature $T$  from an initially ``cooler'' (smaller $T$) sample could possibly
be overtaken by that of an initially ``hotter'' one, if the initial excess kurtosis of the latter is larger enough. We build on
and quantify the implications of this physical idea in the following.

First, we consider the uniformly heated case (i.e., $\chi\neq 0$) and prepare the granular fluid in an initial state that is close
to the steady one, in the sense that Eqs.~\eqref{evol-eqs} can be
linearized around the stationary values \cite{vNE98,MS00}
$T_\st=\left(\chi/\mu_{2}^{\st}\right)^{\frac{2}{3}}$ and
$a_2^\st=[5\mu_2^{(0)}-\mu_4^{(0)}]/[\mu_4^{(1)}-5\mu_2^{(1)}]$, where $\mu_{n}^{\st}=\mu_{n}^{(0)}+\mu_{n}^{(1)}a_{2}^{\st}$.

Let us use a dimensionless
temperature $\theta=T/T_{\st}$ and define $\delta \theta=\theta-1$, $\delta a_{2}=a_{2}-a_{2}^{\st}$, and  $\tau=\kappa \sqrt{T_{\st}}t$.
A straightforward calculation gives
\begin{equation}\label{evol-eqs-linear}
   \frac{d}{d\tau}
\begin{pmatrix} \delta \theta\\ \delta
     a_{2}
\end{pmatrix}=-\mathsf{\Lambda}\cdot
\begin{pmatrix} \delta \theta\\
\delta a_{2}
\end{pmatrix},
\end{equation}
where the matrix $\mathsf{\Lambda}$ has elements
${\Lambda}_{11}=\mu_{2}^{\st}$, ${\Lambda}_{12}= \frac{2}{3}\mu_{2}^{(1)}$,
${\Lambda}_{21}=-2\mu_{2}^{\st}a_{2}^{\st}$,
and
${\Lambda}_{22}=\frac{4}{15}[\mu_{4}^{(1)}-5\mu_{2}^{(1)}(1+a_{2}^{\st})]$.
Thus,  the relaxation of the temperature reads
\begin{eqnarray}
\label{temp-relax-explicit}
\delta\theta&=&\frac{1}{\gamma}\left[(\lambda_{+}-\mu_{2}^{\st})\delta\theta_{0}-
\frac{2}{3}\mu_{2}^{(1)}\delta a_{2,0}\right]e^{-\lambda_{-}\tau}\nn
  &&
-\frac{1}{\gamma}\left[(\lambda_{-}-\mu_{2}^{\st})\delta\theta_{0}-
     \frac{2}{3}\mu_{2}^{(1)} \delta a_{2,0}\right]e^{-\lambda_{+}\tau},
\end{eqnarray}
where $\lambda_\pm=\frac{1}{2}\left[\Lambda_{11}+\Lambda_{22}\pm\sqrt{(\Lambda_{11}-\Lambda_{22})^2+4\Lambda_{12}\Lambda_{21}}\right]$
are the eigenvalues of the matrix $\mathsf{\Lambda}$ and
$\gamma\equiv\lambda_{+}-\lambda_{-}>0$.

\begin{figure}
\includegraphics[width=3.25in]{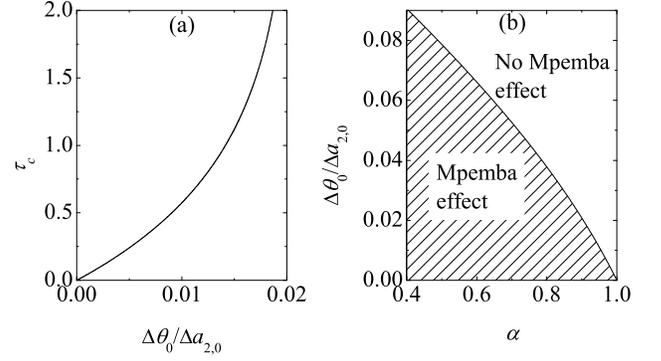}
\caption{\label{fig:deltaT-max} (a) Crossover time $\tau_c$ as a
  function of the ratio $\Delta \theta_{0}/\Delta a_{2,0}$ for
  $\alpha=0.9$. (b) Phase diagram in the plane
  $\Delta \theta_{0}/\Delta a_{2,0}$ vs $\alpha$. The regions of the
  plane inside which there appears or does not appear the ME are
  separated by the curve
  $(\Delta \theta_{0}/\Delta a_{2,0})_{\text{max}}$.}
\end{figure}

Let us imagine two initial states $f_0=f_A$ and $f_B$, with
$(\theta_{0},a_{2,0})=(\theta_{A},a_{2A})$ and $(\theta_{B},a_{2B})$,
respectively. We assume that $\theta_{A}>\theta_{B}$ and
$a_{2A}>a_{2B}$. Both cooling ($\theta_{A}>\theta_{B}>1$) and heating
($\theta_{B}<\theta_{A}<1$) processes may be considered. From
Eq.~\eqref{temp-relax-explicit}, the time $\tau_{c}$ for the possible
crossing of the two relaxation curves satisfies
\begin{equation}\label{crossing-time}
\tau_{c}= \frac{1}{\gamma}\ln
\frac{2\mu_{2}^{(1)}-3(\lambda_{-}-\mu_{2}^{\st})\Delta \theta_{0}/\Delta a_{2,0}}{2\mu_{2}^{(1)}-3(\lambda_{+}-\mu_{2}^{\st})\Delta \theta_{0}/\Delta a_{2,0}},
\end{equation}
where $\Delta \theta_{0}=\theta_{A}-\theta_{B}$ and
$\Delta a_{2,0}=a_{2A}-a_{2B}$.
For a given $\alpha$, in this simplified description the crossover time
$\tau_c$  depends on $(\theta_A,a_{2A})$ and $(\theta_B,a_{2B})$ (or,
more generally, on the details of the two initial VDFs $f_A$ and
$f_B$) \emph{only} through the single control parameter $\Delta
\theta_{0}/\Delta a_{2,0}$.

Figure \ref{fig:deltaT-max}(a) displays $\tau_c$ as a function of the ratio $\Delta \theta_{0}/\Delta a_{2,0}$ for $\alpha=0.9$.
Equation~\eqref{crossing-time} implies that there is a maximum  of the control parameter
$\Delta \theta_{0}/\Delta a_{2,0}$ for which the ME can be observed, namely,
\begin{equation}\label{Delta-max}
\left(\frac{\Delta \theta_{0}}{\Delta a_{2,0}}\right)_{\text{max}}=\frac{2}{3}\frac{\mu_{2}^{(1)}}{\lambda_{+}-\mu_{2}^{\st}}.
\end{equation}
This quantity determines the phase diagram for the occurrence of the ME, as shown in Fig.~\ref{fig:deltaT-max}(b).

Equation~\eqref{Delta-max} can be read in two alternative ways. First,
it means that, for a given difference $\Delta a_{2,0}$ of the initial
kurtosis, the ME appears when the difference $\Delta \theta_{0}$ of
the scaled initial temperatures is below a maximum value
$(\Delta \theta_{0})_{\text{max}}$, proportional to $\Delta a_{2,0}$.
Second,
for a given value of $\Delta \theta_{0}$, the ME is observed only for
a large enough difference of the initial kurtosis, i.e.,
$\Delta a_{2,0}>(\Delta a_{2,0})_{\text{min}}$, with
$(\Delta a_{2,0})_{\text{min}}$ proportional to $\Delta \theta_{0}$.
This quantitatively measures how different the initial conditions of
the system must be in order to have the ME.

\begin{figure}
\includegraphics[width=3.25in]{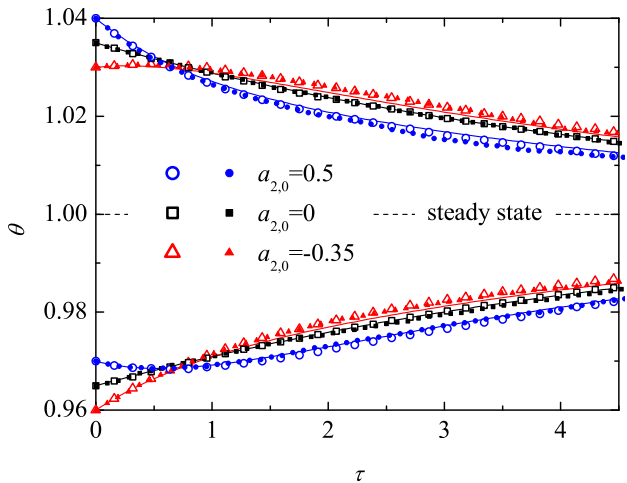}
\caption{\label{fig:relaxation} Relaxation of the scaled temperature
to the steady state for $\alpha=0.9$. The upper and the lower curves
correspond to the ME for the cooling and heating processes,
respectively (see text). The direct simulation Monte Carlo (DSMC)  (open symbols) and molecular dynamics (MD) (filled
  symbols) data show an excellent agreement with the theoretical
  prediction (lines).}
\end{figure}

In order to check the accuracy of our theoretical results, we compare
them in Fig.~\ref{fig:relaxation} with MD simulations (at a density
$n\sigma^3=0.02$) and with the numerical integration of the
Enskog-Fokker-Planck equation by means of the DSMC method \cite{B94}. In all our simulations,
$\alpha=0.9$ and the initial VDF is assumed to have a
gamma-distribution form \cite{HC78} in the variable $v^2$ with
parameters adjusted to reproduce the chosen values of $\theta_{0}$ and
$a_{2,0}$.  First, three different initial conditions (\emph{A}, \emph{B}, and \emph{C})
with temperatures above the stationary, $\theta_{A}=1.04$,
$\theta_{B}=1.035$, and $\theta_{C}=1.03$, and excess kurtosis
$a_{2A}=0.5$, $a_{2B}=0$, and $a_{2C}=-0.35$, are considered. The ME
is clearly observed as a crossover of the relaxation curves of the
temperature [see, also, Fig.\ \ref{fig:deltaT-max}(a)].  Second, we
analyze a ``heating'' protocol by choosing initial temperatures
below the steady value, namely, $\theta_{A}'=0.97$,
$\theta_{B}'=0.965$, and $\theta_{C}'=0.96$, with the same values of
the excess kurtosis as in the ``cooling'' case. Again, a crossover in
the temperature relaxation curves appears, signaling the granular
analog of the \textit{inverse} ME proposed in a recent work
\cite{lu_anomalous_2016}.  It is interesting to note that the
evolution curves corresponding to $\theta_{C}=1.03$ and
$\theta_{A}'=0.97$ are nonmonotonic. This peculiar behavior is
predicted by Eq.~\eqref{temp-relax-explicit} to take place if
$-\frac{2}{3}\mu_2^{(1)}/\mu_{2}^\st<\delta\theta_0/\delta a_{2,0}<0$.

\begin{figure}
\includegraphics[width=3.in]{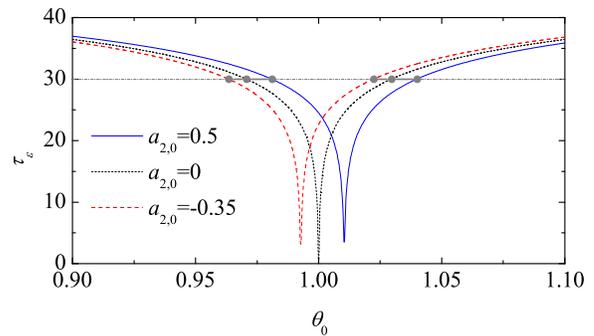}
\caption{\label{fig:relax-time} Relaxation time $\tau_{\epsilon}$ (with $\epsilon=10^{-4}$) as a
  function of the initial scaled temperature $\theta_{0}$ for $\alpha=0.9$. Three values of the initial excess
  kurtosis are considered, $a_{2,0}=0.5$ (solid line), $a_{2,0}=0$
  (dotted line), and $a_{2,0}=-0.35$ (dashed line). The horizontal (grey) segments
  join values of initial temperatures that share the same value of
  the relaxation time and, thus, mark the onset of either the ME
  ($\theta_0>1$) or the inverse ME ($\theta_0<1$).}
\end{figure}

Alternatively, we can characterize the
system celerity for cooling (or heating) by defining a relaxation time $\tau_{\epsilon}$  as the time after
which $|\theta(\tau_{\epsilon})-1|<\epsilon$, with $\epsilon\ll 1$. From Eq.~\eqref{temp-relax-explicit},
\begin{equation}
  \label{relaxation-time}
  \tau_{\epsilon}=\frac{1}{\lambda_{-}} \ln\left|
    \frac{3(\lambda_{+}-\mu_{2}^{\st})\delta
      \theta_{0}-2\mu_{2}^{(1)}\delta a_{2,0}}{3\epsilon\gamma}\right| .
\end{equation}
 Figure \ref{fig:relax-time} shows $\tau_{\epsilon}$ as a function of the
initial temperature $\theta_{0}$ for $\epsilon=10^{-4}$ and the same
values of the initial excess kurtosis as considered in Fig.\
\ref{fig:relaxation}. In this diagram, for a given pair of $a_{2,0}$,
the range of initial temperatures for which the ME emerges is clearly
visualized. Note that this range does not change if the value of the
bound $\epsilon$ is changed to $\epsilon^{\prime}$, since the diagram
is only shifted vertically by an amount $\frac{1}{\lambda_{-}}\ln(\epsilon/\epsilon^{\prime})$.

A relevant question is whether or not the ME keeps appearing in the zero driving limit. In the undriven case ($\chi=0$),
the granular fluid relaxes to the so-called {homogeneous cooling state}
(HCS) \cite{H83}, which is the reference state for deriving the granular
hydrodynamics \cite{BDKS98}. If the linear relaxation picture
developed above remained valid in the nonlinear relaxation regime, at
least qualitatively, the answer would be negative. Note that the
maximum temperature difference $(\Delta T_{0})_{\text{max}}$ would
vanish in the limit as $\chi\to 0$ ($T_{\st}\propto\chi^{2/3}\to 0$),
as a consequence of $(\Delta\theta_{0})_{\text{max}}$ being
independent of $\chi$. Interestingly, we show below that this simple
scenario does not hold, and the ME is also observed for very small
driving: indeed, $(\Delta T_{0})_{\text{max}}$ remains finite in
this limit.

For very small driving, there is a wide initial time region inside which the
system evolves as if it were cooling freely. Therefore, for the sake
of simplicity, we now take $\chi= 0$ in the
evolution equations \eqref{evol-eqs}. While the system freely cools
for all times ($\lim_{t\to\infty}T=0$), the excess kurtosis tends to a constant value \cite{vNE98,MS00}
$a_2^\hcs=[5\mu_2^{(0)}-\mu_4^{(0)}]/[\mu_4^{(1)}-\mu_4^{(0)}-5\mu_2^{(1)}]$.
Since there
is no natural temperature scale in the free cooling case, we can make use of
dimensionless variables by scaling temperature and time with an arbitrary
reference value $T_{\re}$, i.e., $T^{*}=T/T_{\re}$ and $t^*=\kappa \sqrt{T_\re} t$.

If present at all, we expect the ME to occur for relatively
short times, more specifically, before $a_{2}$ has relaxed to its
stationary value $a_{2}^{\hcs}$. So as to look for a possible
crossover of the cooling curves, we linearize the equations around
$T^{*}=1$ (by choosing $T_{\re}$ such
that the initial temperatures verify $|T_{0}^{*}-1|\ll 1$) and
$a_2=a_{2}^{\hcs}$.
 Therefrom, the evolution of $T^{*}$ is
obtained as
\begin{eqnarray}
\delta T^{*}&=&\left(\delta  T_{0}^{*}+\frac{2}{3}-\frac{2}{3}\frac{\mu_{2}^{(1)}\delta
a_{2,0}}{\lambda_{a}-\mu_{2}^{\hcs}}\right)e^{-\mu_{2}^{\hcs}t^{*}} \nn
&&
+\frac{2}{3}\frac{\mu_{2}^{(1)}\delta
a_{2,0}}{\lambda_{a}-\mu_{2}^{\hcs}}e^{-\lambda_{a}t^{*}}-\frac{2}{3}, \label{HCS-temp-evolution}
\end{eqnarray}
where $\delta T^{*}=T^{*}-1$, $\delta {a}_{2,0}=a_{2,0}-a_{2}^{\hcs}$, and
$\lambda_{a}=\frac{4}{15}[\mu_4^{(1)}-5\mu_2^{(1)}-2\mu_4^{(0)}+5\mu_2^{(0)}]$.
In
turn, $a_{2}$ decays exponentially to $a_{2}^{\hcs}$ with a
characteristic time $\lambda_{a}^{-1}$.

\begin{figure}
\includegraphics[width=3.25in]{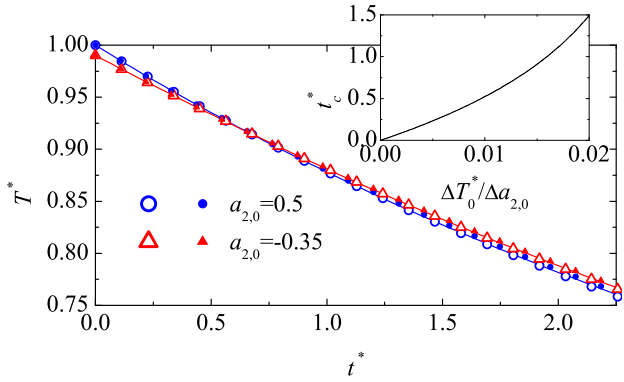}
\caption{\label{fig:HCS-relaxation} Evolution of the temperature in
  the free cooling case.   Again, the agreement between the theory (lines) and both  DSMC  (open symbols) and MD (filled symbols)  simulation data  is excellent. The inset shows $t_c^*$ as a function of $\Delta T_{0}^{*}/\Delta a_{2,0}$.}
\end{figure}

Similar to the thermostatted case, we consider two initial
states $(T_{0}^{*},a_{2,0})=(T_{A}^{*},a_{2A})$ and
$(T_{B}^{*},a_{2B})$, with $\Delta T^{*}_0=T_{A}^{*}-T_{B}^{*}>0$,
$\Delta a_{2,0}=a_{2A}-a_{2B}>0$. Logically, only the cooling case makes
sense. In Fig.~\ref{fig:HCS-relaxation}, we plot two relaxation curves
of the temperature for $\alpha=0.9$, with $T_{A}^{*}=1$,
$T_{B}^{*}=0.99$, $a_{2A}=0.5$, $a_{2B}=-0.35$, with the choice
$T_{\re}=T_{A}$.  The ME is clearly observed, and
the crossover time $t_{c}^{*}$  is
\begin{equation}\label{crossing-time-cooling}
t_{c}^{*}=\frac{1}{\lambda_{a}-\mu_{2}^{\hcs}}
\ln\left(1-\frac{3}{2}\frac{\lambda_{a}-\mu_{2}^{\hcs}}{
  \mu_{2}^{(1)}}\frac{\Delta T^{*}_0}{\Delta a_{2,0}}\right)^{-1};
\end{equation}
see inset in Fig.~\ref{fig:HCS-relaxation}. Therefore, there is a
  maximum value of the ratio $\Delta T^{*}_0/\Delta a_{2.0}$ for which
  the ME appears,
\begin{equation}\label{deltaT-max-cooling}
  \left(\frac{\Delta T^{*}_0}{\Delta a_{2,0}}\right)_{\text{max}}=\frac{2}{3}\frac{\mu_{2}^{(1)}}{\lambda_{a}-\mu_{2}^{\hcs}}.
\end{equation}

Thus, the ME actually survives in the zero driving limit.  Had we
considered a small value of the driving $\chi$ instead of $\chi= 0$,
Eqs.~\eqref{HCS-temp-evolution}--\eqref{deltaT-max-cooling} would
characterize the strongly nonlinear regime, in which the initial
scaled temperature $\theta_{0}=T_{0}/T_{\st}\gg 1$. In a first stage
of the relaxation, as long as the granular temperature $T\gg T_{s}$,
the driving can be neglected, the system freely cools, and the ME is
observed provided that the condition \eqref{deltaT-max-cooling} is
fulfilled. Afterwards, the initially hotter system remains below the
initially cooler one forever. When approaching the steady state, both
the temperature and the excess kurtosis start to evolve towards their
stationary values $T_\st$ and $a_2^\st$, but in both curves one has
$a_{2,0}=a_{2}^{\hcs}$, and Eq.~\eqref{Delta-max} tells us that no
further crossing of the curves takes place ($\Delta a_{2,0}=0$).


In summary, we have shown by means of a simple analytical theory that the ME
naturally appears in granular fluids, as a consequence of the relevance
of non-Gaussianities in the time evolution of $T$. Specifically, this
allows us to (i) prove that the ME is to be expected on quite a general
basis and for a wide range of systems, as long as non-Gaussianities are present and (ii) quantitatively predict
the region of parameters within which the ME is present. Moreover, we have
also predicted the existence of an \textit{inverse} ME: when the
system is heated instead of cooled, the initially cooler sample may
heat sooner \cite{lu_anomalous_2016}. In this way, we have provided a
general theoretical framework for the understanding of the ME.

The main assumptions in our theory are: (i) the validity of the
kinetic description, (ii) the system remaining homogeneous for all
times, and (iii) the first Sonine approximation within the kinetic
description. All these assumptions have been validated in the
paper. First, the numerical integration of the Enskog equation
provided by the DSMC simulations  has been successfully compared with MD
simulations. Second, we have also checked that
the system remains homogeneous in the MD simulations, both for the
heated and undriven cases. Concretely, in the latter, the system size
has been chosen to be well below the clustering instability threshold
\cite{GZ93,AT09}. Third, the accuracy of the first Sonine
approximation has been confirmed by the excellent agreement between
our analytical results and the DSMC simulations, even for the
not-so-small values of the excess kurtosis $a_{2}$ considered
throughout.

Finally, we stress that non-Gaussianities may have a leading role in
the emergence of the ME in other systems, even when there is no
inelasticity. For example, the temperature of a molecular fluid, which
is basically the mean kinetic energy per particle, does not remain
constant if the system interacts with a thermal reservoir.  Let us
assume that the coupling with the reservoir brings about a
\emph{nonlinear} drag, as considered, for instance, in
Refs.~\cite{klimontovich_nonlinear_1994,SAG95,F98,M13}. Then, the evolution equation of the
temperature would involve higher moments of the transient
nonequilibrium VDF. In this quite general situation, the ME would also
stem from those non-Gaussianities \cite{LVPySunpub}.

\begin{acknowledgments}
  This work has been supported by the Spanish Ministerio de Econom\'ia y
  Competitividad Grants No.~FIS2013-42840-P (A. L., F. V. R., and
  A. S.), No.~FIS2016-76359-P (F. V. R. and A. S.), No.~MTM2014-56948-C2-2-P (A. L.), and No.\ FIS2014-53808-P (A. P.), and also by the Junta de
  Extremadura Grant No.~GR15104, partially financed by the European Regional Development Fund
  (A. L., F. V. R., and A. S.). Computing facilities from Extremadura
  Research Centre for Advanced Technologies (CETA-CIEMAT) funded by
  the European Regional Development Fund are also acknowledged. A. P. thanks M. I. Garc\'ia de Soria
  for very helpful discussions.
\end{acknowledgments}

\bibliography{Mpemba_granular,Granular}

\end{document}